\newcommand{\beq}{\begin{quote}}
\newcommand{\enq}{\end{quote}}
\newcommand{\be}{\begin{equation}}
\newcommand{\en}{\end{equation}}
\newcommand{\del}{\delta}

\documentclass[11pt]{amsart}
\usepackage{geometry}                
\usepackage{graphicx}
\usepackage{amssymb}
\usepackage{epstopdf}
\title{Visiting Newton's Atelier Before the {\it Principia}, 1679-1684.}
\author{Michael Nauenberg\\
University of California Santa Cruz}

\begin{document}
\begin{abstract}
 
 The manuscripts   that presumably contained  Newton's early development of the fundamental concepts that  led to his {\it Principia}
have  been lost.  A plausible   reconstruction of  this development  is presented  based on Newton's   
exchange of letters with Robert Hooke in 1679, with Edmund Halley in 1686, and on some clues in the diagram associated with
 Proposition1 in Book1 of  the {\it  Principia}  that have been ignored in the past.  The graphical method associated with this proposition  leads to a rapidly convergent method  to obtain orbital curves for central forces,  and elucidates how Newton may have have been led to  formulate some of his other propositions  in the {\it Principia}.     
. 
\end{abstract}

\maketitle
\section{Introduction}
The publication of Newton's masterpiece, ``Mathematical Principles of Natural Philosophy"  known as  {\it Principia} \cite{cohen1}, marks  the beginning  of  modern {\it theoretical} Physics and  Astronomy. It was 
 regarded as a very difficult book  by his contemporaries,  and also by modern readers. In 1687 when the {\it Principia} was first published,  it 
was  claimed that only a  handful of  readers in Europe were competent  to read it \cite{hall}.  John Locke, for example,  found the demonstrations impenetrable, and asked Christiaan Huygens if he could trust them.  When Huygens assured him that he could, ``he applied himself to the prose and digested the physics without the mathematics" \cite{westfall}. Shortly  after the release
 of the {\it Principia}   a group of students at Cambridge supposedly were heard by Newton  to say, ``here goes a man who has written a book
 that neither he nor anyone else understands" \cite{quote}. Newton himself remarked that 
 `` to avoid being baited by little Smatterers in Mathematicks . . . he designedly made his {\it Principia} abstruse;  but yet as to be understood  by able Mathematicians" \cite{snobolen}

Presumably one person he had in mind  was Robert Hooke, with whom he had often quarreled. But shortly after viewing an early draft of the {\it Principia}, entitled  {\it De Motu Corporum  Gyrum}, that Newton had sent to the Royal Society in 1685 \cite{flamsteed},
Hooke  understood  that Newton had implemented
mathematically his own  concept of orbital dynamics that he had communicated to Newton in a correspondence in 1679.  Hooke  promptly applied the first theorem in {\it De Motu}, and  in an unpublished manuscript dated Sept 1685,  he  showed  graphically and analytically  that for a central  attractive force depending {\it linearly} on the distance from the center, the resulting orbit is an ellipse \cite{pugliese},\cite{michael4}. Newton described his concepts in a geometrical language  rather than in the analytic calculus that he and Gottfried Leibniz had developed.   At the time  that the {\it Principia} was first  published (1687) this approach made sense,  because  the calculus, first  published by Leibniz in 1684, was hardly known, and Newton did  not published his own version (Fluxions) in full until 1704.  It has been  argued that Newton's  geometrical descriptions  were also the way in which he originally discovered and developed his fundamental dynamical concepts \cite{hall2}. This view will  be  supported  here. It is also claimed that the {\it Principia} is replete with mathematical abstractions and incomprehensible geometrical diagrams  \cite{cohen3}. Even Richard Feynman remarked that he could not follow Newton's  demonstration in Proposition 11 ({\it Principia}, Book 1) that elliptical orbits  imply an inverse square  central force \cite{ feynman}. But  these difficulties are often overemphasized. It will be shown that   with an elementary knowledge of geometry,  Proposition 1, one of  the most fundamental  theorems on which the  {\it Principia} is based,  can be  applied as a {\it graphical method} to construct approximate orbits that describe  the dynamical effect of central forces.

 There is considerable evidence that a major progress in Newton's  understanding of orbital 
dynamics occurred  in 1679  through  his correspondence with  Robert Hooke, who described to 
Newton his  concept for  the physical basis of 
 planetary motion. Hooke viewed   the origin of the  motion of planets around the Sun  by  the  {\it compounding}  of inertial  motion with {\it periodic } gravitational impacts towards the Sun. This concept  was  the  link missing  before  Newton could develop his earlier  dynamical concepts further \cite{michael}. In a letter to Newton on  Nov. 24,1679, Hooke wrote,
 ``For my own part I shall take it as a great favour if you please to communicate by Letter your 
 objections against any hypothesis  or opinion of mine, And particularly if you will let me know your thoughts of
 that  of compounding  the celestiall motion of the planetts of a direct motion by the tangents and an attractive motion toward the central body" \cite{hooke1}. 
 
 Hooke had elaborated his dynamical ideas in a short tract, published in 1674,   entitled ``An attempt to prove the motion the motion of the Earth by observations" \cite{hooke4}. He argued that attractive  gravitational forces were universal,  and regarding terrestrial  gravitation he wrote:
 ``This propagated Pulse I take it to be the Cause of the descent of bodies towards the Earth . . . Suppose for  Instance  there should be
 1000 of these Pulses in a second of Time,  then must the Grave body receive  all those thousand impressions within the space of time
 of that Second, and a thousand more the next . . ." \cite{hooke2}, \cite{galileo}. In his letter to Newton  he did not specify that the gravitational force
 could be regarded  as a sequence of impacts, but there is evidence that Newton was familiar with Hooke's 1674 tract.  In  a letter to Halley on July 14, 1686, Newton admitted that: 
``This is true, that his Letters occasioned my finding the method of determining Figures which when I tried in the Ellipsis, 
I threw the calculation by being upon other studies \& so it rested for about 5 years till upon your request I sought for yt
paper, \& not finding it did it again and reduced it into ye  Proposition shown you for Mr. Paget . . ."\cite{newton1} 

Newton's new ``method  of determining Figures"  appeared in the first Theorem of {\it De Motu} and in  Proposition of the {\it Principia}, Book 1,  where it is presented, however,  in the formal  mathematical language of a theorem and its  proof,  instead of as a graphical method to obtain  orbital curves under the action of  a central attractive force.
This theorem as it appeared in {\it De Motu}  states that``All orbiting bodies describe, by radii drawn to their centre,  areas proportional to their times" \cite{whiteside1}. 
  For planetary motion, this relation, known {\it empirically} by astronomers as the ``area law",  was  originally conjectured  by Kepler from the observations of planetary  motion by Tycho Brahe, but without any dynamical understanding of its origin. For the special case of a circular orbit transversed with constant velocity,  the area law follows from the relation between  the circumference and the  area of the circle,  shown  first by Archimedes. In effect, Newton's geometrical construction  is a generalization of  Archimedes' construction  that  approximates a general curve by an inscribed polygon subdivided into  triangles  of equal areas with a common center.

The simplest case of Newton's  geometrical construction  of  orbital curves by the action of periodic central force impacts  
  is  for constant impacts.  As Newton wrote to Halley on May 27, 1686, ``the simplest case for computation,  which was that of Gravity uniform in a medium not Resisting" \cite{halley}.
  This remark, however,  has puzzled  historians of science  aware that  a constant central attractive force does not
 have a simple analytic  solution \cite{newton8} But to obtain  such a discrete  orbit   for periodic impacts {\it graphically} ,
 all that is needed is a {\it ruler and a pencil}. Indeed,  an examination of Newton's diagram for Theorem 1 in {\it De Motu}
 and for Proposition 1 in the {\it Principia}, shows that both  were  carefully drawn to {\it scale},   for the special case of {\it constant impacts}. In the resulting diagram in  {\it De Motu},  reproduced in Fig.1,   and  in a somewhat different diagram  in the {\it Principia},  he presented  only the discrete orbit resulting   for the first four impacts, which was  perfectly adequate as an illustration
 for  his  theorem or  proposition \cite{chatelet}.  But it would be  highly unlikely  that he would not have followed his drawing with  further steps as he had  done earlier for the case of a constant central force  by his curvature method \cite{michael},  and compare the  results from his two distinct
 methods (see Fig. 4, panels A and B).
 
  When discussing the continuum limit of infinitesimal triangles in Proposition 1,  Newton referred  to Lemma 3, Corollary 4,
   indicating  that the 
lines representing the  impacts in his diagram ended  on a given  curve representing the continuum orbit.  But this curve does
not appear in Newton's diagram.  Moreover,  for finite impacts the
length of these impact lines would not scale with the  radial dependence under consideration. It will be shown that to keep the local radius of curvature of the continuous 
orbital curve  constant,  the magnitude of the impact lines vary  {\it quadratically} relative  to the magnitude of the  inertial lines (Section 2, Eq. 7).  In this case the convergence  of the resulting discrete polygon to a  continuous orbital curve occurs rapidly.  

A  bonus from  the  graphical method to obtain a  polygonal approximation for  the orbital curve  resulting from central force impacts is that  it also gives the time elapsed to reach
 each vertex of the polygon, This time  is  proportional to the number of vertices  from the initial vertex of the orbit.
Then the approximate velocity at each vertex is obtained by dividing  the average displacement from  two adjacent vertices 
by the periodic  time between impacts. In the {\it Principia},  the orbital curve and elapsed  time can be obtained
 ``granting the quadrature of curvilinear  figures",  meaning, by 
evaluating  the  two integrals given in  in Proposition 41, Problem 28, which in general has
to be done  numerically.  There is not even a hint that it can also be obtained 
by a graphical method.

 Following Newton's  graphical construction,  in Section 2  orbital curves  are obtained  with the initial variables scaled according to the diagrams in {\it De Motu},  and in the {\it Principia}.  For the case of constant impacts, the results are  in  good agreement with Newton's earlier approximate graphical calculation based on his curvature method \cite{michael} - see Fig.4, panels A amd B - and with an
 experiment of a ball rolling in an  inverted cone - see  Fig. 4 panels  C and D -  carried out also by Hooke \cite{michael4}.
Hooke's  graphical  construction for the case that the impacts vary linearly with the distance
from a fixed center is shown in Fig.6, panel A.  But with the same initial conditions, I found that
 the corresponding  graphical construction for impacts 
that vary inversely with the square of the distance from the center, shown in Fig.6, panel B,  fail to converge  \cite{michael1} .
   Section 3 contains a discussion of the convergence of the the
 discrete orbital curves to the continuum limit, and Section 4 contains a summary and  some conclusions.

   \begin{figure}[htbp] 
   \centering
   \includegraphics[width=5in]{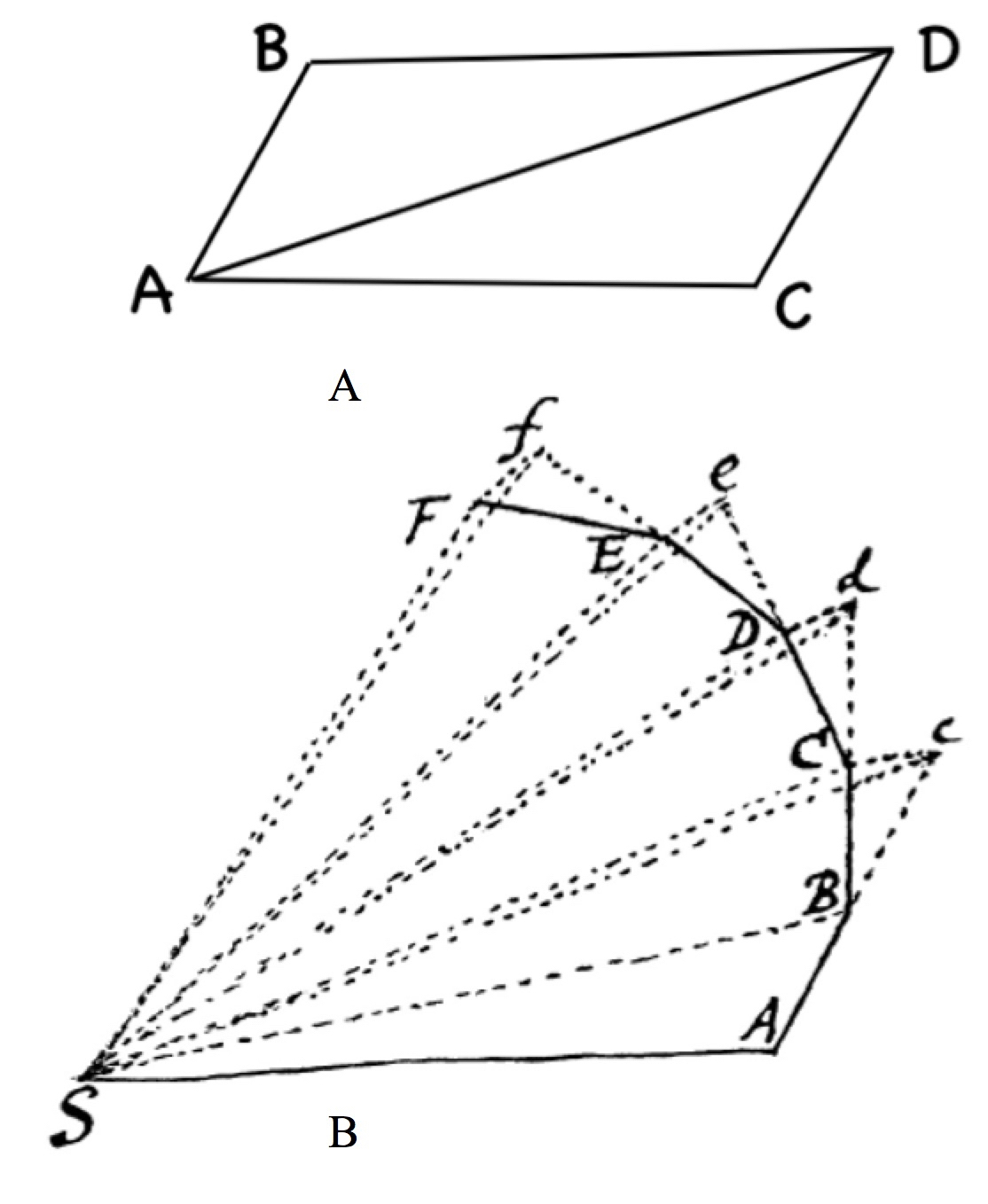} 
   \caption{A.  Diagram for the composition of two velocities.  B. Diagram for Theorem 1 in {De Motu}
   }
\end{figure}
\clearpage

\section {Graphical method to construct orbittal curves by  Proposition 1}

Newton's recipe  to construct graphically a discrete orbit under the action of periodic impacts  as described in  Theorem 1 of {\it  De Motu}, and later in Proposition 1, {\it Principia}, Book 1,  is  as  follows: 
 
Referring to Fig.1, panel B, for  the diagram that appeared in {\it De Motu},   SA  is  the initial distance of a body at A from the center of force at S,  and AB is the straight line  it transverses  at a constant velocity $v$  during an interval of time $\del t$. When it reaches the point B the body receives an impact towards S corresponding in the continuum limit  to an acceleration  $a$,   changing instantaneously  its  velocity by an amount  $\del  v=a \del t$ in this direction.   During the next time interval,  if the body had  started  from rest,  it would transverse the distance $\del v \del t$ corresponding to $Cc$  in the diagram. But the actual distance transversed  and its direction is BC, obtained  by {\it compounding} (adding vectorially) these two velocities. This law of compounding velocities is shown
 in Fig.1, panel A.   for  two simultaneous impacts in the directions AB and AC leading to the displacement AD.  
The line BC is obtained graphically by extending the line AB to c with Bc=AB, drawing Cc from c parallel to SB,
and joining B to C. 
 
This graphical operation is then repeated  {\it successively}  leading to the result shown for constant impacts 
in the diagram for Theorem 1 in {\it De Motu} , Fig.1, panel B obtaining  the  vertices C,D,E and F. 
An illustration how and additional step is obtained is shown in  Fig.2, panels A,B,C,D and the result for 11 additional
inpacts is shown in panel E.

For convenience in notation, let   $AB=d$ and $cC=h$. Then  
\be
\label{vel}
d=v\del t,
\en
and  setting $\del v$ for  the magnitude of the  velocity change due to the component $h_t$ of the impact {\it tangential} to the local motion
\be
\label{acc}
h_t=\del v \del t,
\en
  which can be expressed 
 by a quantity $a_t$  with the dimension of acceleration, 
\be
\del v= a_t \del t.
\en
 Substituting this expression in Eq.\ref{acc}, 
\be
h_t=a_t\del t^2.
\en
To express $h_t$ graphically as a length,  the time interval $\del t$ in this relation  is replaced by $d/v$, Eq.\ref{vel}, and
\be
h_t=\frac{a_t d^2}{v^2}
\en
Finally, replacing $v^2 $  by the product of $ a_t$ times a length $\rho$ corresponding  the local   radius of curvature of the 
continuous orbit at B,
\be
\label{curvature}
\rho=\frac{v^2}{a},
\en
and we have
\be
\label{quadratic}
h= \frac{d^2}{\rho}.
\en

The relation given in Eq.\ref{quadratic},  indicating that $h$ depends quadratically on $d$,
is very important to obtain  improved discrete approximations for  the orbit in the
continuum limit.  For example, if the initial magnitude of  $d$  
is decreased by a factor $1/2$, $h$ should be  decreased  by a factor $1/4$,
and the number of steps required to reach a comparable point of the discrete orbit is approximately doubled.
Although the  value of $\rho$ is not known {\it a priori},  it is not required for the graphical calculation, but 
it can be evaluated approximately at a given vertex as the  radius of a circle that contains this vertex and the
two adjacent ones. Newton discussed the radius of curvature of orbital curves   in Proposition 6, and in Lemma 11 in the {\it Principia}, Book 1.

Given the initial displacement $d$,  the  initial value of $h$ depends on the magnitude of the  impact and can be determined by Eq. \ref{quadratic} if the local curvature $\rho$ of the orbit
at A is  known.  In the diagram in Fig. 1,  which evidently Newton drew carefully and  to scale, he chose the
displacement  $cC=h$ approximately
$1/5$ the length of AB=d.  Subsequent  values of $h$, e.g. $ Dd,Ee,Ff$,  are  determined  by the graphical construction and  the dependence of the magnitude of the impacts on the distances  $SC,SD,SE,SF$ of the vertices from center of force  at $S$. For example, assuming a power law dependence of the impacts on distance with exponent
$p$, the impact $h_c$  at $C$ is related to the previous impact $h_b$  at B by the relation   $h_c=h_b(SC/SB)^p. $, and more generally,
\be 
\label{impact} 
  h_f=h_b(SF/SB)^p
\en
for the magnitude of the impact at any vertex F.
For impacts independent of distance, $p=0$, for a linear dependence on distance, $p=1$,  and for an inverse square dependence $p=-2$. Newton's diagram in {\it de Motu}, shown in Fig. 1,  and the corresponding diagram in {\it Principia} were carefully drawn to scale for {\it constant} impacts, $p=0$. This  is the ``simplest case" for this graphical computation as Newton pointed out  to Halley in 1685 (see the quotation in the Introduction), because it avoids the need of an algebraic computation of Eq.\ref{impact} at each vertex of the discrete orbit .

Regardless of the dependence of the impacts on distance, Newton gave an elementary proof that the areas of successive  triangles 
$SAB,SBC,SCD,SDE,SEF$ obtained by joining the ends of the straight lines  $AB,BC,CD,DE,EF$  between impacts to the center at S
are all equal. Since by construction,  these lines are path lengths transversed at equal time intervals $\del t$ and constant velocity,   the areas enclosed by the  corresponding polygonal path and the center of force are proportional to the time. This is Newton's  proof of the area law theorem in Proposition 1 quoted in the Introduction.

 \begin{figure}[htbp] 
   \centeringÝ
   \includegraphics[width=6in]{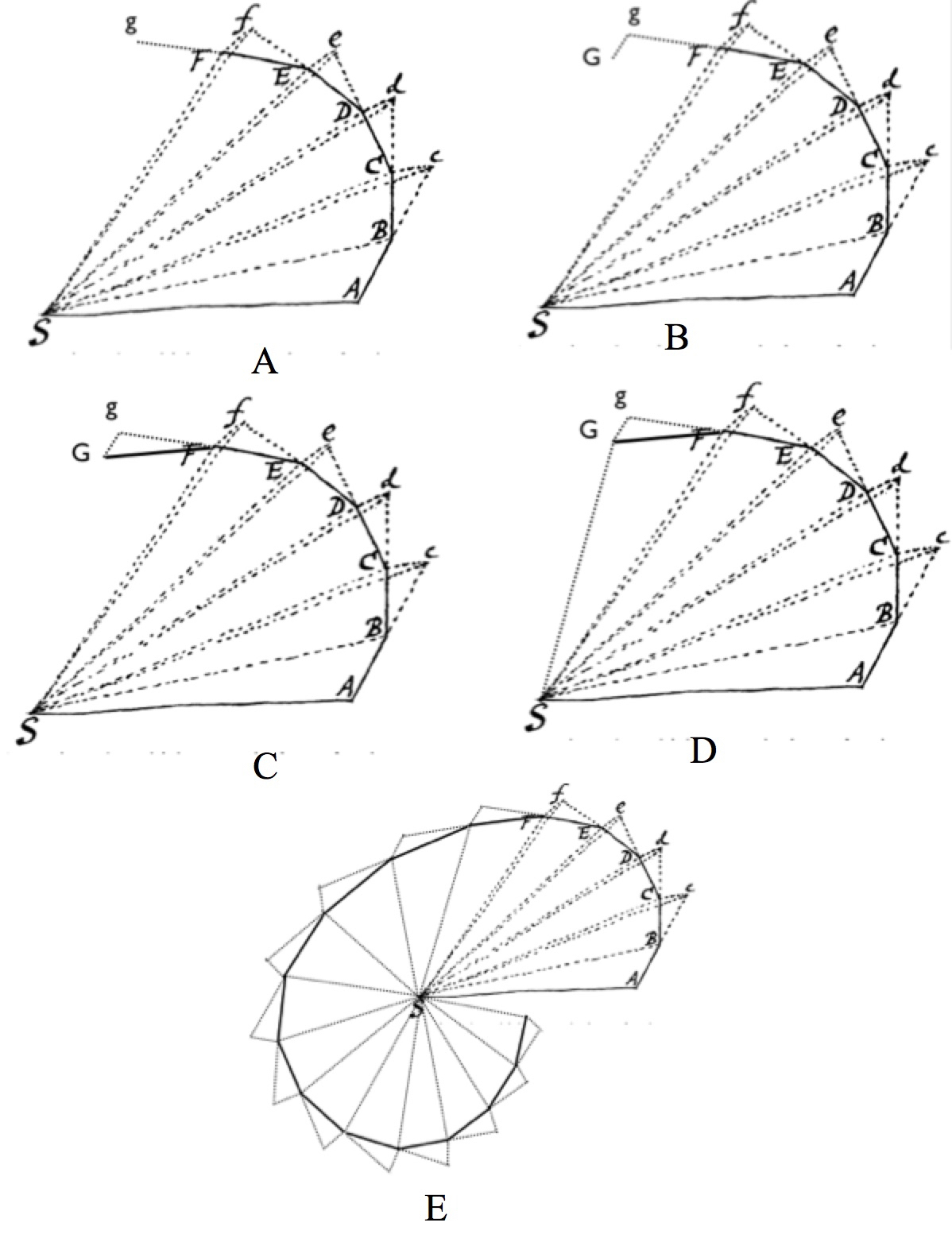} 
   \caption{Graphical extension of Newton's  diagram for Theorem 1 in {de Motu} . Panels A,B,C and D show the
   successive graphical operations  to add a single step to Newton's diagramm.  Panel E is  the resultant  orbit after adding 12 impacts.
   }
\end{figure}
\clearpage

 \begin{figure}[htbp] 
   \centeringÝ
   \includegraphics[width=6in]{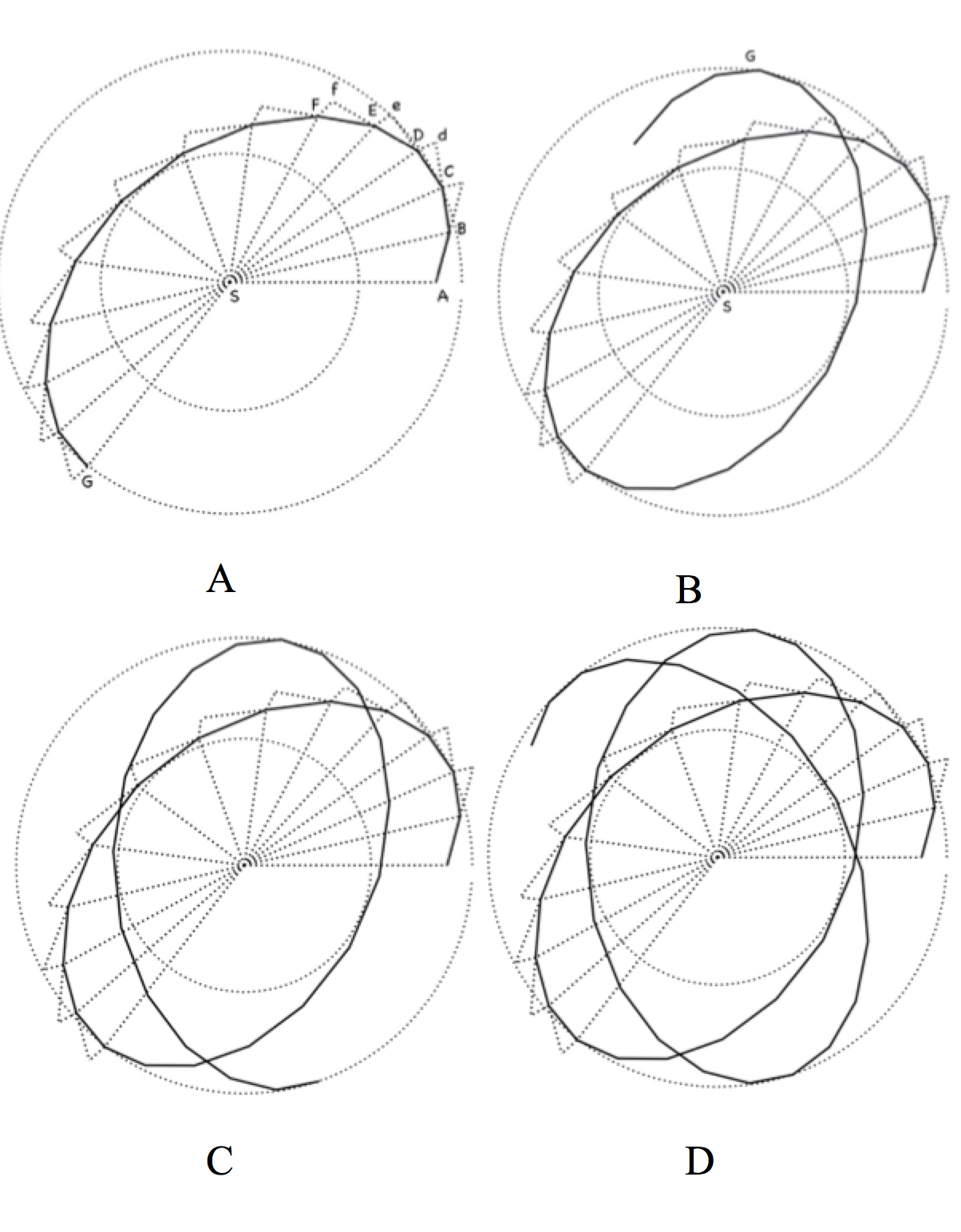} 
   \caption{Graphical extension of Newton's  diagram for Proposition 1 in {\it Principia},  demonstrating
   the confinement of the orbit between two circles. Panel B,  obtained by doubling the
   number of steps in panel A, illustrates also a  reflection symmetry property around the axis GS of the orbit in panel A.
   Panels C and D illustrates  further branches of the orbit.
   }
\end{figure}
\clearpage

Although Newton's  diagram in Prop.1 was drawn to  scale for only  4  equal impacts,   it is very likely  that he would  have
continued this graphical construction for more steps as shown in Figs 2 and 3.   In Fig. 2,  panels A,B,C and D  illustrates how  the extension of one triangle at the end  of vertex F of this diagram is obtained:
Taking  E as the  initial starting point, panel A illustrates the first step, drawing the line Fg equal in length and direction to EF,
 panel B   the impact length Gg drawn equal in length to the other impacts and parallel to SF,
panel C   the  line joining G an F for the displacement GF,   and panel D   the line SG for the distance and direction at G. 
Panel E  shows the resulting discrete
orbit for 11 more impacts. It is very likely that  Newton would  have carried out this  graphical  extension, but 
there isn't  any manuscript evidence for it \cite {cohen5}.

 \begin{figure}[htbp] 
   \centering
   \includegraphics[width=6in]{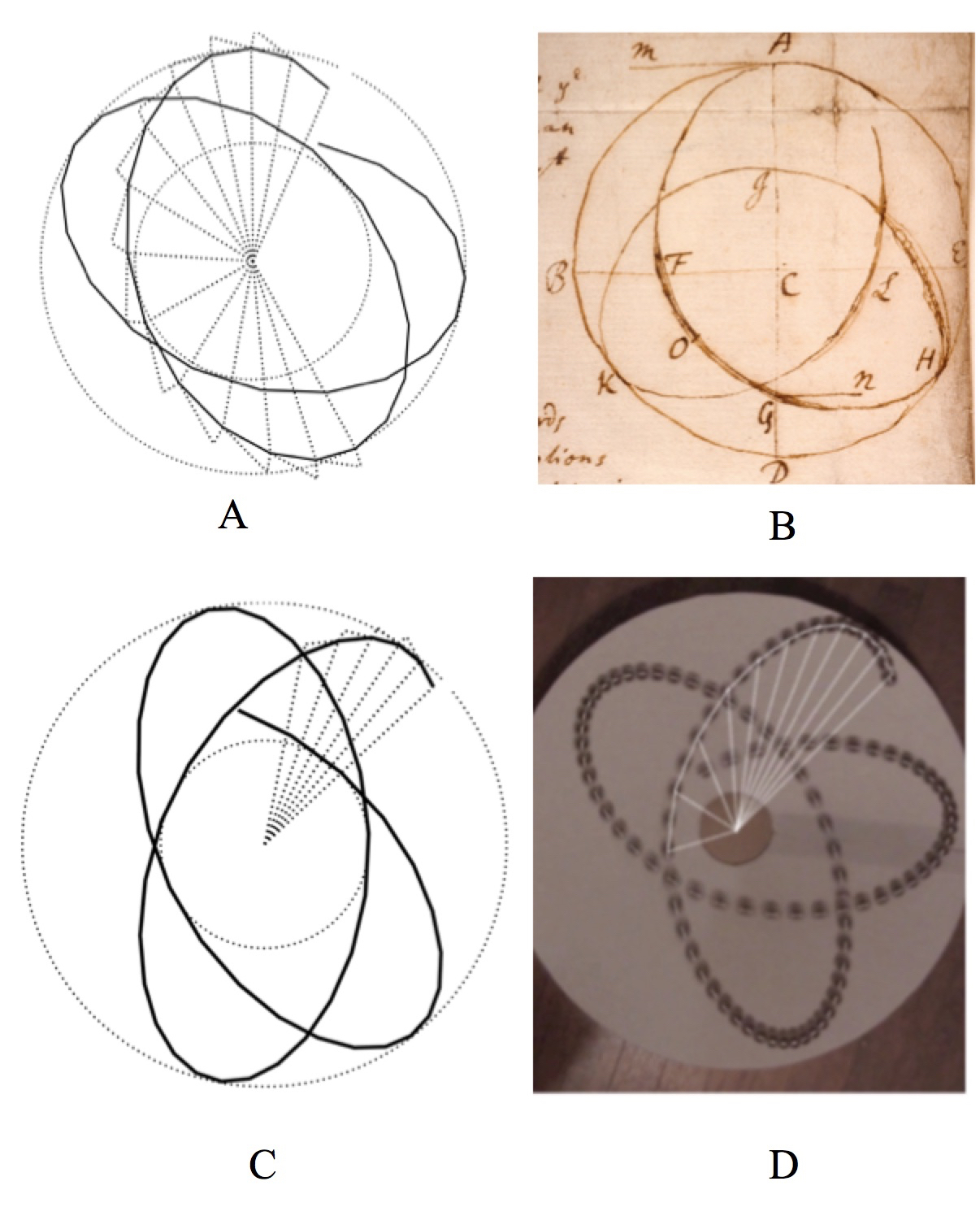} 
   \caption{  A.  Extension of Proposition 1 diagram for constant impacts. 
B.  Newton's diagram in a letter to Hooke on Dec. 13, 1679.  C. Extension of De Motu
diagram with impacts adjusted to fit the  experiment shown in  the next panel.
   D. Stroboscopic view of a ball rolling in an inverted cone
   } 
   \end{figure}
\clearpage

 Newton's graphical construction for constant impacts is continued for additional impacts, and the results are shown in Fig. 3, panels A, B, C, and D. These  diagrams show that the orbit has  the  shape  of an ellipse with  its  axis  rotating clockwise about the center of force (impacts),
 Panel B  shows that the next branch of this orbit can be obtained by a reflection about the axis GS indicated in panel A, and similarly
 in panels C and D. This orbit  is confined between  two circles with radii corresponding to the nearest and farthest  distance of the
 orbit from the center of force.  This restriction is due to the time reversal property of this graphical construction that will not be discussed
here.

 Fig. 4, panel A   shows the similarity of the discrete orbital curve  obtained with constant impacts, and  the diagram in panel B of  a figure that Newton included in a letter to Hooke, dated Dec.13, 1679. Newton sent Hooke  the diagram shown in   
this panel , without describing how he had obtained  it except for a  short  comment that ``I here consider motion according to the method of indivisibles" \cite{newton7}. Hooke  promptly  responded that ``Your calculation of the Curve by a body attracted by an equall power at all 
Distance from the Center Such as a ball Rouling  in an inverted  Concave Cone
is right, and the two auges will not unite  for about a third of a Revolution " \cite{hooke3}, \cite{hooke3b}.
Evidently this experiment had been carried out  by Hooke who became very excited after learning 
that Newton had  been able to calculate the corresponding orbit  by an  appproximate graphical method \cite{michael1}.
At the  time Newton could not have  implemented the impact 
 approach, because,  by  his own admission, he had developed it  only after his  correspondence with Hooke.
 An alternative approach to obtain  the diagram in panel B was based   on the local curvature of an orbital curve
 also developed independently by Christiaan Huygens \cite{michael}. 
  Panel D shows a stroboscopic view of an experiment \cite{michael11} with a ball
rolling inside an inverted cone.  In this case  the radial force is approximately constant, and the radial lines for the first 11  consecutive position have been drawn  to illustrate  Kepler's  Area Law (conservation of angular momentum). Panel C shows the orbital curve obtained
graphically for constant impacts with similar initial conditions.  The differences from the experimental orbit  is partly due to friction which
is evident by the decrease in the maximum displacement  of this trajectory from its center.

   \begin{figure}[htbp] 
   \centering
   \includegraphics[width=6in]{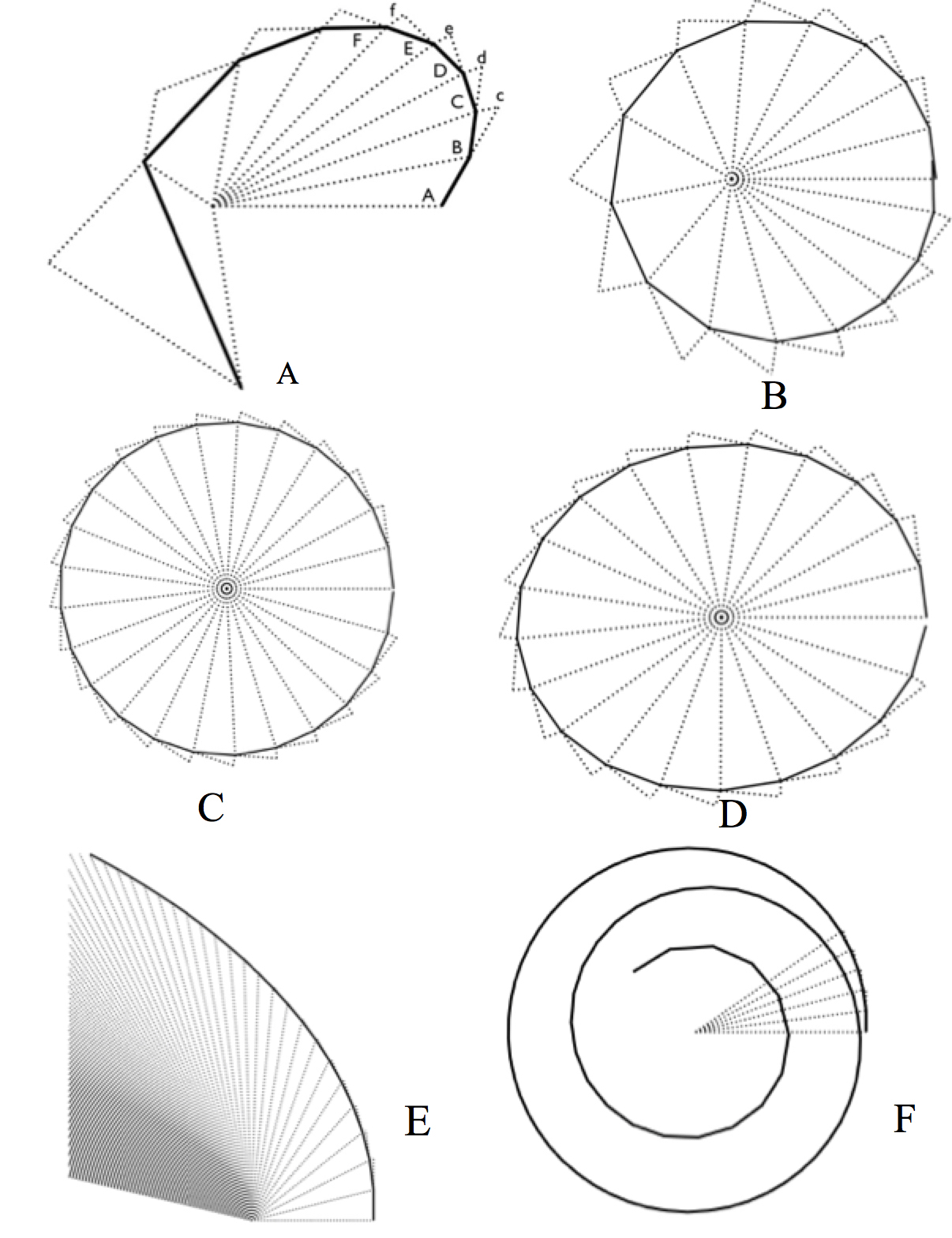}
   \caption{ A. Extension of  the  Proposition 1 diagram for  inverse square impacts, p=2. B. Elliptical orbit  for p=2.
C. Crcular orbit for p=0. D. Elliptical orbit  for p=-1. E. Hyperbolic orbit for p=2. F. Spiral orbit
for p=3
    }
\end{figure}
\clearpage

 Fig. 5, panel A,  shows the resulting discrete orbit when the impacts are scaled according to
 an inverse square force, starting with the same initial conditions for  constant impacts in the {\it De Motu}
 diagram. After  the  eight impact, the orbit gets too close to the center of force and starts to diverge.  Panel B
 shows a discrete elliptical orbit with the center of force at a focus of the ellipse. It is obtained  when the initial displacement 
 is inclined at an angle  that gives an isosceles triangle in the first step of the graphical construction. 
  For  congruent isosceles triangles associated with the impact length  the  resulting orbit
  the graphical construction gives a circular orbit shown in panel C .  Panel D shows a discrete  elliptical orbit  for impacts  linear dependent
  on the distance from the center of the ellipse. Panel  E shows a discrete hyperbolic orbit for inverse
  square force.  The envelope of the radial lines at the bottom of this figure  indicates the direction
  of its  asymptote.  Panel F shows an inward spiraling orbit 
  resulting  for the inverse cube force.  A the end
  of his Dec. 13, 1679 letter to Hooke,  Newton remarked ``for the increased of gravity in the descent may be
  supposed such that the body shall by an infinite number of spiral revolutions descend continually  till it cross
  the  center by motion transcendentally swift"\cite {newton7}. Evidently,  at the time Newton was already aware that
  for an inverse cube  force the orbits do not remain confined. Later, in Proposition 9 in the {\it Principia},  he gave a
  proof that for a spiral orbit, ``the centripetal force is inversely as the cube of the distance (from the center) SP"\cite{cohen2}.

 Fig. 6, panel A, shows the upper right hand part of a diagram in a manuscript  by Hooke, dated Sept.1685,
two years before the publication of the {\it Principia}.  It represents  a discrete
 orbit obtained with Newton's graphical procedure,  for a body rotating clockwise under the action of a sequence of impacts
towards the center at O, that depend linearly on the distance to this center. The resulting vertices of this discrete orbit lie on an ellipse with the center of force at the center of the ellipse. An analytic  proof of this result was also given by Hooke\cite{michael1}.  Panel B is the  diagram I obtained for an inverse square force  with initial conditions equal to those in Hooke's 
diagram. For the first 7 steps the discrete orbit approximates closely  to an ellipse, as  shown by dots in this  diagram, with a focus  at F,  the center of force.  But after the discrete orbit approaches  closely  to F,  in the next two steps the impact lengths increase too
rapidly as was the case with Newton's orbit shown in Fig. 5, panel A, and the graphical construction diverges. It is likely that
Hooke also carried out this graphical calculation, and  perhaps discouraged by this divergence,  he failed to publish his results.

 \begin{figure}[htbp] 
   \centering
   \includegraphics[width=6in]{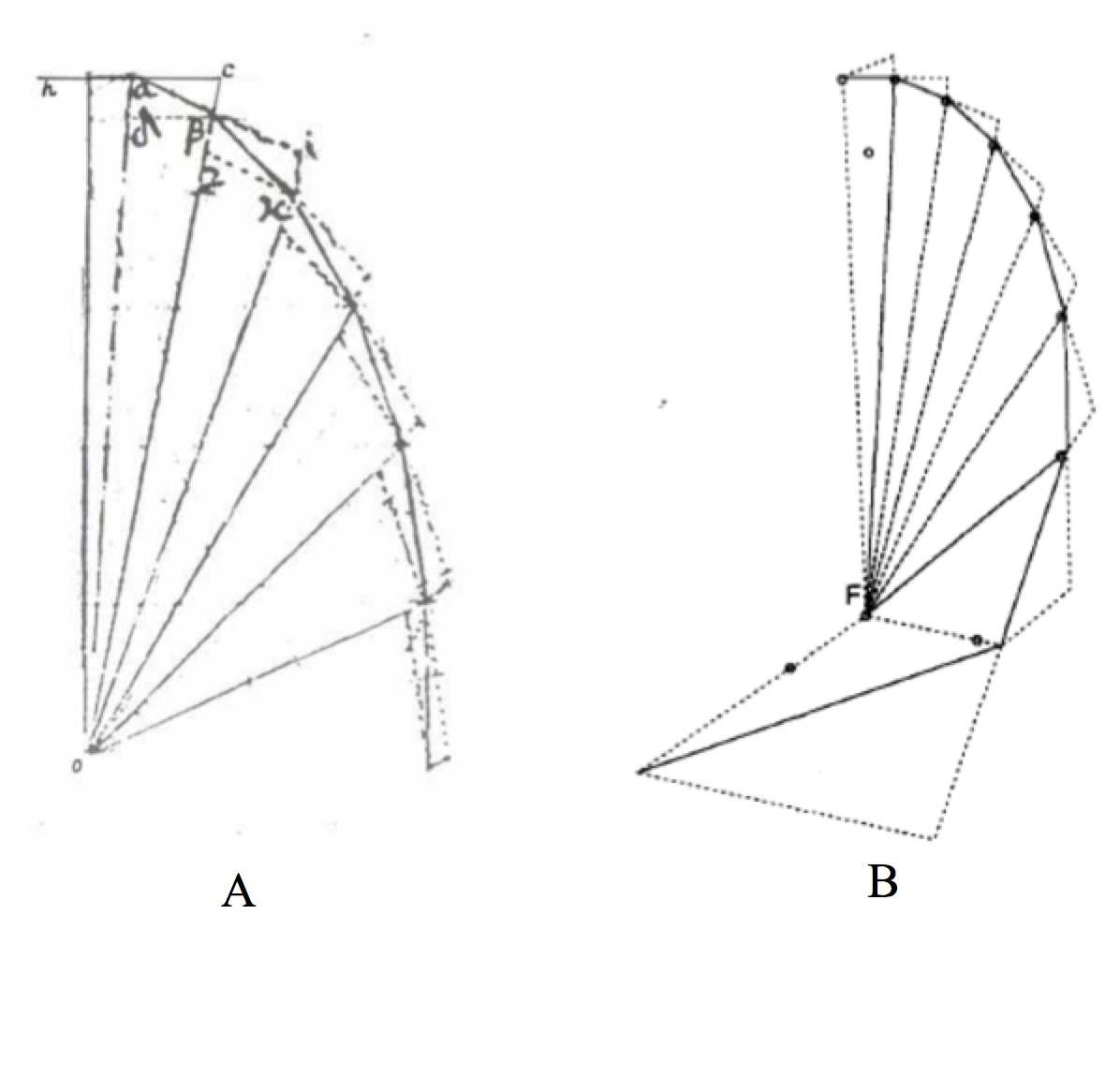}  
   \caption{ A. Upper right hand part of HookeÕs Sept.1685 diagram for a discrete
elliptical orbit rotating clockwise under the action of a sequence of impulses
towards O, that depend linearly on the distance to this center.  B. Corresponding diagram
for an inverse square force  with initial conditions similar to those in Hooke's 
diagram. }
\end{figure}

\clearpage

\section{Summary and Conclusions}
It has been shown that Newton's geometric proof of Kepler's Area Law in Theorem 1 in  {\it De Motu},  and  in  Proposition 1 of  the  {\it Principia}, Book 1,  also  describes  a  graphical method to obtain   approximate orbits under the action of  
central forces. When Newton sent  {\it De Motu} to the Royal Society in 1684,  Hooke, who was at the time the secretary of the RS, 
 obtained the  copy\cite{flamsteed}, and recognized that Newton had implemented
geometrically his concept of planetary motion -  compounding inertial motion with a gravitational  attraction towards the Sun.
A manuscript dated Sept 1685 in the collection of his papers in  the Trinity College  Library in Cambridge, shows that Hooke promptly  applied Newton's graphical
construction  to obtain a discrete orbit  under the action of central force  impacts that
depend {\it  linearly} on the distance from a fixed center (see Fig. 6, panel A) \cite{pugliese}, \cite{michael4}. He  also gave an analytic proof that the vertices of the resulting polygonal  orbit  lie on an ellipse with the center of the attractive force located at the center of this ellipse.  He had observed this orbit  with a pendulum, but I could not find  any  evidence among his papers in this library that he also tried to evaluate the discrete orbit resulting  for the gravitational  inverse square force.  Had he carried it out with his previous initial conditions, he would  have found that the  graphical procedure  fails  when the  vertices of the discrete orbit approach too closely to  the center of force, as shown in my graphical calculation in Fig. 6,  panel B. The same problem occurs also  if one starts  with Newton's initial conditions,  as  shown in Fig.5,  panel A.  It is plausible that Hooke  carried out this calculation, but could not figure out  the source of this problem, and for this reason did not   publish   his results.  
 
On Nov 24, 1679 Hooke communicated to Newton his  physical concept for the orbital motion 
 under the gravitational action of a central force.  It appears to be very unlikely that after Newton  implemented 
 Hooke's concept  geometrically, that he would not apply his construction, as Hooke did in 1685,  to obtain such  orbits graphically. In particular for the {\it simplest} case of constant impacts, Newton's initial condition in the diagrams associate with Theorem 1 in {\it De Motu} and with  Proposition 1 in the {\it Principia}, leads to an approximate discrete orbit  shown in Fig. 2,  panel  E, and in Fig. 3, panel A. This orbit is in good agreement with the result that he had shown to Hooke on Dec 13, 1679, (see  Fig 3, panel C)  which  I conjectured was obtained by a different graphical method based
 on the concept of curvature developed independently  by Huygens\cite{michael}. In {\it De Motu} and in the {\it Principia}, Newton' diagrams are  shown only for four impacts  which was  adequate to illustrate his proof of  Kepler's Area Law, see Fig.1 panel B.  In both cases these diagrams were  drawn carefully  to  scale for the case of constant impacts, and one expects that Newton would   have been interested in comparing the resulting discrete orbit with further impacts shown in Fig.4, panel A ,
 with his earlier calculation  shown Fig. 4, panel  B..  This  orbit looks like an ellipse with its  axis rotating uniformly,  and most
 likely was also the  inspiration for his  remarkable  Proposition 45, in the {\it Principia}, Book1, that states: 
 `` It is required to find the motion of the apsides of orbits that differ very little from circles" \cite{cohen2}.
 
 When Halley wrote to Newton that Hooke wanted his contributions to be acknowledged in the {\it Principia},  Newton angrily replied:
 `` For  tis plain by his words he knew not how to go about it.  Now is this not very fine? Mathematicians that find out, settle \& 
 do all the business must consent themselves with being nothing but dry  calculators \& drudges \& another that does nothing but
 pretend \& grasps at all things must carry away all the invention . . .\cite{newton10}. During a visit at Halley's,  Hooke met Newton and reported in his diary``vainly pretended claim yet acknowledge my information, Interest has no conscience "\cite {hooke11}.

 Had Newton chosen to describe Proposition 1 in the {\it Principia} not only as a mathematical  proof of 
 Kepler's area law,  but also as  a graphical method to calculate approximately 
 orbits for central forces,  and had he  also given for illustration  some examples  as those  shown here, it would have made
 his book  accessible to a large number of readers familiar with elementary geometry.  But he chose otherwise, presumably
 ``to avoid being baited by little Smatterers in Mathematicks "\cite{snobolen}. Moreover, I have shown that  the  initial lines  in his  diagrams for Theorem 1 in {\it De Motu},  and in  Proposition 1 in {\it Principia}  lead  after only a few steps to   a divergent orbit for impacts that vary inversely with the square of the distance from the center.(see Fig 5, panel A). Were Newton's   choices of  initial parameters 
 in his diagram for Proposition 1 made  to discourage  little Smatterers in mathematics?
Hooke also found such a  divergence with his own choice of initial conditions (see Fig. 6 panel B), and presumably became discouraged from pursuing his graphical constructions further\cite{michael4}, \cite{michael5}.
 
 In his introduction to Newton's {\it Principia}, the eminent Newtonian scholar  I. B. Cohen asks:
 ``Whatever happened to the work-sheets of the {\it Principia}? Do they still exist in some obscure private or public collection?
 Was this particular set of manuscripts - alone of all the Newton papers - lost or mislaid, either when the  Portsmouth Collection
 was still in Hurstbourne Castle or during the actual transfer to the  University Library in Cambridge? Did such work-sheets still exist
 among Newton's papers at the time of his death? Or were they lost or destroyed - either by chance or design - during Newton's 
 own lifetime? We may possibly never be certain of the answer to these questions."\cite{cohen6} \cite{dry}.
 
 In the absence of these work-sheets a plausible reconstruction has been given here,  based on the first  two fundamental propositions on which the {\it Principia} is based: Propositions 1 and 6. To underscore the importance of  such preliminary work we conclude by
 quoting Simon Laplace's: ``knowledge of the method that has guided a man of genius is no less  useful to the progress of science and to his glory than his discoveries; the method is often the most interesting part".

\section{Acknowledgement}
I dedicate this article to the memory of Bruce Brackenridge with whom I had many fruitful discussions on Newton's {\it Principia}.   
I thank Niccol\'{o} Guicciardini and David Book  for their helpful comments.


\begin{thebibliography}{30}
\bibitem{flamsteed}  On December 27, 1684,  Flamsteed wrote to Newton:  ``  . . . how much I am obliged by your kind concession
of ye perusall of your papers, tho I believe I shall not get a sight of them till our common friend  Mr. Hooke \& the rest of the town have
been first satisfied".  Reference 3, p. 405.
\bibitem{cohen3}  For example, I. B. Cohen wrote that `` the issue of the mathematical rigor in Newton's analysis has been and still
	                      remains, a subjectd of debate among scholars. It has been argued, for example (by Michel Blay among others), that
	                      even if the limit of the polygon can legitimately be taken to be  a continuous curve, the limit of a series of impulsive
	                      blows cannot be a continually acting force".  Reference 1,  p. 71, foonote 73.
\end{thebibliography}
\end{document}